\documentstyle[12pt,epsf]{article}
\makeatletter
\@addtoreset{equation}{section}
\makeatother
\def\xib{\bar{\xi}}

\begin{document}

\begin{titlepage}

\def\thefootnote{\fnsymbol{footnote}}
\setlength{\baselineskip}{13pt}

\begin{flushleft}
BONN-TH-95-06\hfill March 1995\\
ITP-UH-10/95\hfill cond-mat/9503142\\
ITP-SB-94-71
\end{flushleft}

\vspace*{\fill}

\begin{center}
{\large\bf Integro-Difference Equation for a correlation function\\
        of the spin-${1\over2}$ Heisenberg XXZ chain}
\vfill

{\sc Fabian H.L. E\char'31ler}\footnote{%
e-mail: {\tt fabman@avzw02.physik.uni-bonn.de}}\\

{\sl Physikalisches Institut der Universit\"at Bonn\\
     Nussallee 12\\
     D--53115~Bonn, Germany}\\[8pt]
{\sc Holger Frahm}\footnote{%
e-mail: {\tt frahm@itp.uni-hannover.de}}\\
{\sl Institut f\"ur Theoretische Physik\\
     Universit\"at Hannover\\
     D--30167~Hannover, Germany}\\[8pt]
{\sc Alexander R.\ Its}\footnote{%
e-mail: {\tt itsa@math.iupui.edu}}\\
{\sl Department of Mathematical Sciences\\
     Indiana University-Purdue University at Indianapolis (IUPUI)\\
     Indianapolis, IN 46202--3216, U. S. A.}\\[8pt]
{\sc Vladimir E.\ Korepin}\footnote{%
e-mail: {\tt korepin@max.physics.sunysb.edu}}\\
{\sl Institute for Theoretical Physics\\
     State University of New York at Stony Brook\\
     Stony Brook, NY 11794--3840, U. S. A.}\\[8pt]
\vfill
ABSTRACT
\end{center}

\begin{quote}
We consider the {\em Ferromagnetic-String-Formation-Probability}
correlation function (FSFP) for the spin-$1\over 2$ Heisenberg XXZ
chain. We construct a completely integrable system of integro-difference
equations (IDE), which has the FSFP as a $\tau$-function. We derive the
associated Riemann-Hilbert problem and obtain the large distance
asymptotics of the FSFP correlator in some limiting cases.
\end{quote}

\vfill

\setcounter{footnote}{0}
\end{titlepage}

\section{Introduction}

In this paper we continue our investigation of a particular
zero-temperature correlation function of the XXZ Heisenberg model in the
critical regime $-1<\Delta<1$ in an external magnetic field.
The XXZ hamiltonian is given by
\begin{equation}
   {\cal H} = \sum_{j} \sigma_j^x \sigma_{j+1}^x
                     + \sigma_j^y \sigma_{j+1}^y
                     + \Delta\ (\sigma_j^z \sigma_{j+1}^z-1)
                     - h \sigma_j^z\ ,
   \label{xxx}
\end{equation}
where the sum is over all integers $j$, $\sigma^\alpha$ are Pauli matrices
and $h$ is an external magnetic field. For later convenience we define
$\Delta = \cos(2\eta) = {w +{1\over w}\over 2}$, where ${\pi\over 2}<\eta<
\pi$. The FSFP correlation function is defined as follows
\begin{equation}
P(m) = \langle 0 |\; \prod_{j=1}^m P_j \;| 0 \rangle ,
   \label{px}
\end{equation}
where $|0\rangle$ is the antiferromagnetic ground state and $P_j={1\over2}
(\sigma_j^z + 1)$ is the projection operator onto the state with spin up at
site number $j$. The physical meaning of $P(m)$ is the probability of
finding a ferromagnetic string (i.e.\ $m$ (adjacent) parallel spins up) in
the ground state $|0\rangle$ of the model (\ref{xxx}) for a given value of
the magnetic field $h$. {}From a mathematical point of view this correlator
turns out the simplest one to be considered (see \cite{vladb}).

In a recent paper \cite{efik:94} we have derived a representation of the
correlation function (\ref{px}) as a determinant of a Fredholm integral
operator
\begin{equation}
   P(m) = { { \left( 0 \left|
                  \det \left(1 + {\widehat{V}^{(m)}} \right)
                       \right| 0 \right) }\over
            { \det \left( 1 - {1\over 2\pi}\widehat{K} \right) }}\ .
   \label{fred}
\end{equation}
Here $\widehat{K}$ and $\widehat{V}^{(m)}$ are integral operators
acting on a function $f(\lambda)$ as
\begin{equation}
   \left( \widehat{V}^{(m)} f \right) (\lambda)
          = \int_{-\Lambda}^{\Lambda} d\mu\ V^{(m)}(\lambda|\mu)
f(\mu)\ ,\quad
   \left( \widehat{K} f \right) (\lambda)
          = \int_{-\Lambda}^{\Lambda} d\mu\ K(\lambda|\mu) f(\mu)\ ,
\label{intop}
\end{equation}
where $\Lambda$ is a function of the anisotropy $\Delta$ and the magnetic
field $h$.  The kernels are given by (as compared to \cite{efik:94} we have
performed a similarity transformation, which leaves
$\det(1+\widehat{V}^{(m)})$ invariant but changes the kernel
$V^{(m)}(\lambda|\mu)$)
\begin{equation}
   V^{(m)}(\lambda|\mu) = -{1\over 2\pi}
      \frac{\sin(2\eta)}{\sinh(\lambda -\mu)}
     \left({e^{(m)}_+(\lambda)e^{(m)}_-(\mu)
              \over (\sinh(\lambda - \mu +2i\eta)} -
           {e^{(m)}_-(\lambda) e^{(m)}_+(\mu)
              \over \sinh(\mu - \lambda +2i\eta)}
     \right)\ ,
   \label{vkernell}
\end{equation}
\begin{equation}
   K(\lambda|\mu) = {\sin(4\eta)\over
                    \sinh(\lambda-\mu+2i\eta)\sinh(\lambda-\mu-2i\eta)}
\end{equation}
with functions $e^{(m)}_\pm(\lambda) = \left(\left({\sinh(\lambda
+i\eta)\over\sinh(\lambda-i\eta)}\right)^m
e^{\phi(\lambda)}\right)^{\pm{1\over2}}$.  Here $\phi(\lambda)$ is a {\em
dual quantum field} acting in a bosonic Fock space with vacuum $|0)$. The
fields $\phi$ are related to canonical Bose fields $a$ and $a^\dagger$
(annihilating the vacuum and its dual respectively, {\em i.e.}\
$a(\lambda)|0)=0=(0|a^\dagger(\lambda)$) via the Bogoliubov transformation
\begin{equation}
   \phi (\lambda) = a(\lambda)
     - \int^\infty_{-\infty} d\nu
       \ln \bigg[h(\lambda,\nu)h(\nu,\lambda)\bigg] a^\dagger(\nu)\ ,
\label{dualfield}
\end{equation}
where $h(\lambda,\nu)= {\sinh(\lambda-\nu+2i\eta)\over i\sin(2\eta)}$. The
occurrence of the dual fields is a consequence of the interacting nature of
the Heisenberg model (see Section XI.1 of \cite{vladb}). By construction
the fields $\phi(\lambda)$ commute for all values of spectral parameter
$\left[\phi(\lambda)  ,\phi(\mu) \right] = 0$. The contribution of the
dual fields (to the expectation value $(0|\det(1+{\hat V}^{(m)})|0)$)
can be visualized by decomposing all dual fields $\phi$ according to
(\ref{dualfield}) and then moving all exponentials of annihilation
operators $a(\l)$ to the right, picking up contributions whenever
passing by an $a^\dagger(\nu)$.

The purpose of the current paper is to derive a system of
integro-difference equations that drive the Fredholm determinant of the
operator $\widehat{V}^{(m)}$ in the correlation function $P(m)$
(\ref{fred}). This is done in Section~\ref{sec:ide} and can be considered
as an extension of the idea of describing quantum correlation function by
means of differential equations [Painlev\'e] due to E. Barouch, B.M. McCoy,
T.T. Wu and C. Tracy and B.M. McCoy \cite{barouch:73,trmc:73}.  The
Fredholm determinant $\det\left(1-{1\over 2\pi}\widehat{K}\right)$ in the
denominator of (\ref{fred}) amounts merely to an overall normalization
independent of the distance $m$ and will not be considered here. In
Section~\ref{sec:RH-problem} the Riemann-Hilbert problem associated with
the IDE is formulated. Finally, in Section~\ref{sec:special}, we determine
the long-distance asymptotics of the FSFP in some limiting cases.

\section{Integro-Difference Equations\label{sec:ide}}

We start by bringing the kernel of $\widehat{V}$ (\ref{vkernell}) to
``standard'' form \cite{vladb}. We first perform a change of variables $z=
{e^{2\lambda}w-1\over e^{2\lambda}-w}$ (recall that $w=e^{2i\eta}$), which
maps the real axis on the contour $C: \alpha\to z=\exp{i\alpha}$ (see
Fig.~\ref{fig:contour}) where $-\psi < \alpha <2\pi+\psi$ ($\psi<0$
by definition).  The endpoints $\xi = e^{i\psi}$ and ${\xib} = e^{-i\psi}$
(we integrate from $\xib$ to $\xi$) of the contour are related to the
magnetic field $h$ {\em and} the anisotropy $\Delta$. Using the identity
(valid for ${\pi\over 2}<\eta<\pi$ and $z_1,z_2\in C$)
\begin{equation}
   \int_0^\infty ds\ e^{-i(w {wz_1-1\over z_1-w}
                        - {1\over w}{wz_2-1\over z_2-w})s}
   = {-i\over w {wz_1-1\over z_1-w} -{1\over w}{wz_2-1\over z_2-w}}\ ,
\end{equation}
the kernel of $\widehat V^{(m)}$ is found to be (up to a similarity
transform which leaves the determinant unchanged)
\begin{equation}
   V^{(m)}(z_1|z_2) = -{i\over 2\pi}\int_0^\infty ds\
             {e^{(m)}_+(z_1|s) e^{(m)}_-(z_2|s)
            - e^{(m)}_-(z_1|s) e^{(m)}_+(z_2|s)
             \over z_1-z_2}\ .
   \label{vkernelz}
\end{equation}
where the functions $e_\pm^{(m)}$ are given by
\begin{eqnarray}
   e_+^{(m)}(z|s) &=&
     \left( i(w-1/w)\ {wz-1\over z-w}\ z^{-m}e^{-\phi(z)}\right)^{1\over2}
     \exp\left( {i\over w}\ {wz-1\over z-w}\ s\right)\nonumber \\
   \label{def:e} \\
   e_-^{(m)}(z|s) &=&
     \left( i(w-1/w)\ {wz-1\over z-w}\ z^{m}e^{\phi(z)}\right)^{1\over2}
     \exp\left( -iw\ {wz-1\over z-w}\ s\right)\nonumber
\label{e}
\end{eqnarray}
The integral operator $\widehat{V}^{(m)}$ now acts on a function
$f(z)$ as
\begin{equation}
   \left( \widehat{V}^{(m)} f \right) (z_1)
                = \int_C dz_2 V^{(m)}(z_1|z_2) f(z_2)
\end{equation}
where the integration is to be performed along the contour
$C$. We note that $V$ is symmetric and nonsingular at $z_1=z_2$.
The resolvent $\widehat{R}^{(m)}$ of $\widehat{V}^{(m)}$ is defined by
$(1+\widehat{V}^{(m)})(1-\widehat{R}^{(m)})=1$
and its kernel $R^{(m)}(z_1|z_2)$ can be written in a form
similar to Eq.~(\ref{vkernelz}), namely \cite{vladb}
\begin{equation}
   R^{(m)}(z_1|z_2) = -{i\over{2\pi}}\int_0^\infty ds\ {{f^{(m)}_+(z_1|s)
f^{(m)}_-(z_2|s) - f^{(m)}_-(z_1|s) f^{(m)}_+(z_2|s)} \over {z_1-z_2}} \; .
  \label{rkernelz}
\end{equation}
Here $f_\pm^{(m)}$ are solutions of the linear integral
equations $\left( (1+\widehat{V}^{(m)})f_\pm^{(m)} \right) (z|s) =
e_\pm^{(m)}(z|s)$.
In terms of these functions we introduce the integral operators
$B_{ab}^{(m)}$, $a,b=\pm$ acting as $(B^{(m)}_{ab} f^{(m)})(s) =
\int_0^\infty dt\ B^{(m)}_{ab}(s,t) f(t)$ with the kernel
\begin{equation}
  B^{(m)}_{ab}(s,t) = {i\over2\pi} \int_C {dz\over z} f^{(m)}_a(z|s)
e_b^{(m)}(z|t)\, , \qquad a,b = \pm \, .
  \label{def:B}
\end{equation}
The {\em transpose} $B^T$ acts like $\left(B^T f\right)(s) = \int_0^\infty
dt\ B(t,s) f(t)$. We are now in the position to formulate the main

\noindent{\bf Theorem :}
{\em
\begin{itemize}
\item[(i)]
The lattice logarithmic derivative of $\det\left( 1+\widehat{V}^{(m)}
\right)$ is given in terms of the integral operator $B^{(m)}_{ab}$ as
\begin{equation}
   { {\det\left( 1+ \widehat{V}^{(m+1)} \right)} \over
     {\det\left( 1+ \widehat{V}^{(m)} \right)} }
   = \det\left( 1 + B_{-+}^{(m)} \right)
   \label{vvd1}
\end{equation}
\item[(ii)]
The logarithmic derivative of $\det\left(1+\widehat{V}^{(m)}\right)$ with
respect to the boundaries of the contour $C$ is expressed in terms of the
functions $F_\pm^{(m)}(s)=f_\pm^{(m)}(\xi|s)$ and
$G_\pm^{(m)}(s)=f_\pm^{(m)}({\bar\xi}|s)$ as follows
\begin{eqnarray}
   -i \partial_\psi \ln \det\left( 1+\widehat{V} \right) &=& {1\over 2\pi}
     \int_0^\infty ds\ \Bigl\{F_+(s) \partial_\psi F_-(s)
                 - F_-(s) \partial_\psi F_+(s)
       \nonumber \\
  &&\qquad \qquad
                 + G_-(s) \partial_\psi G_+(s)
                 - G_+(s) \partial_\psi G_-(s)
            \Bigr\} \nonumber \\
  && + {1\over 4\pi^2} \; {\xi+\xib \over \xi-\xib} \;
        \left( \int_0^\infty ds\ \left[ F_+(s) G_-(s)
                    - F_-(s) G_+(s)\right]\right)^2 \; .
  \label{eq:vvdpsi}
\end{eqnarray}
\item[(iii)]
The following set of completely integrable integro-difference equations for
the unknowns $F,G,B$ in (i) and (ii) holds
\begin{eqnarray}
   {1 \over \sqrt{\xi}} F_+^{(m+1)} &=& {1\over \xi}
        \left\{ F_+^{(m)}
          - \left( 1- B_{+-}^{(m+1)} \right)
            \left( B_{++}^{(m)} \right)^T F_-^{(m)} \right\}
   \nonumber \\
   {1 \over \sqrt{\xi}} F_-^{(m+1)} &=&
      {1\over \xi} \left\{
        \left( \xi + B_{--}^{(m+1)} \left(B_{++}^{(m)}\right)^T
        \right) F_-^{(m)}
        - B_{--}^{(m+1)} \left( 1+ \left(B_{-+}^{(m)}\right)^T
                         \right) F_+^{(m)} \right\}
   \nonumber \\
   \label{result:f} \\
   {1 \over \sqrt{\xib}} G_+^{(m+1)} &=& {1\over \xib}
        \left\{ G_+^{(m)}
          - \left( 1- B_{+-}^{(m+1)} \right)
            \left( B_{++}^{(m)} \right)^T G_-^{(m)} \right\}
   \nonumber \\
   {1 \over \sqrt{\xib}} G_-^{(m+1)} &=&
      {1\over \xib} \left\{
        \left( \xib + B_{--}^{(m+1)} \left(B_{++}^{(m)}\right)^T
        \right) G_-^{(m)}
        - B_{--}^{(m+1)} \left( 1+ \left(B_{-+}^{(m)}\right)^T
                         \right) G_+^{(m)} \right\}
   \nonumber
\end{eqnarray}
and
\begin{eqnarray}
   -i {\partial \over \partial \psi} B_{ab} (s,t) &=&
    {i\over 2\pi} \Biggl\{ 
      F_a(s) \left[ F_b(t)
                      + \left( F_+ B_{-b} - F_- B_{+b} \right) (t)
                 \right] \nonumber \\
   && \qquad +
      G_a(s) \left[ G_b(t)
                      + \left( G_+ B_{-b} - G_- B_{+b} \right) (t)
                 \right] \Biggr\}
   \label{B:dpsi}
\end{eqnarray}
where $a,b=\pm$.
\end{itemize}

The additional restrictions necessary to solve the equations uniquely are
provided by the requirements on analyticity and the asymptotic behaviour of
the solutions of the corresponding linear system (\ref{shift:fright}),
(\ref{f:dpsi2}), {\em i.e.} by specification of the data in the
corresponding Riemann-Hilbert problem (see Section \ref{sec:RH-problem}).}
\vspace{0.5cm}

\noindent
{\bf Proof:} The proof is analogous to the one for the XXX-case
\cite{frik:94} so that we only sketch the main steps.
(i) is a direct consequence of the {\em shift-equation}
\begin{equation}
   z_1^{-{1\over2}}\; V^{(m+1)}(z_1|z_2)\; z_2^{1\over2}
      = V^{(m)}(z_1|z_2)
      + {i\over2\pi}\int_0^\infty ds\ {e_+^{(m)}(z_1|s)
e_-^{(m)}(z_2|s) \over z_1}\ . \label{shift:v}
\end{equation}
which follows directly from (\ref{vkernelz}) and (\ref{def:e}).
(ii) and (iii) follow from the Lax-representation
\begin{eqnarray}
   {1\over\sqrt{z}} f_-^{(m+1)}(z|s) &=&
       f_-^{(m)}(z|s) - {1\over z} B_{--}^{(m+1)} \left(
             \left[1+ \left(B_{-+}^{(m)}\right)^T\right] f_+^{(m)}
           - \left(B_{++}^{(m)}\right)^T f_-^{(m)} \right)(z|s)\ ,
   \nonumber \\
   {1\over\sqrt{z}} f_+^{(m+1)}(z|s) &=&
       {1\over z} \left(
        f_+^{(m)} - \left[ 1-B_{+-}^{(m+1)}\right]
                    \left(B_{++}^{(m)}\right)^T f_-^{(m)} \right)(z|s)
   \label{shift:fright}
\end{eqnarray}
and
\begin{eqnarray}
   {\partial \over \partial \psi} f_\pm(z|s) &=&
     {1\over 2\pi}\, \left\{ {\xi\over z-\xi} f_\pm(\xi|s)
              \int_0^\infty dt\ \left(f_+(\xi|t) f_-(z|t)
                       - f_-(\xi|t)f_+(z|t)\right)
                    \right. \nonumber \\
  && \quad \left.
      + {\xib\over z-\xib} f_\pm(\xib|s)
              \int_0^\infty dt\ \left(f_+(\xib|t) f_-(z|t)
                       - f_-(\xib|t)f_+(z|t)\right) \right\}\ .
  \label{f:dpsi2}
\end{eqnarray}
{\bf q.e.d} \\
It is a remarkable fact that (\ref{vvd1})-(\ref{B:dpsi}) are formally
identical to the corresponding expressions for the XXX case \cite{frik:94}.

\section{The operator-valued Riemann-Hilbert problem
         \label{sec:RH-problem}}
In this section we show that all results of Section \ref{sec:ide} can be
reformulated in terms of an infinite-dimensional Riemann-Hilbert
problem (RHP) for the integral operator valued function $\chi(z)$.
{}From the solution $\chi(z)$ of this RHP one can then determine an
asymptotic expansion of the correlator $P(m)$.

\begin{figure}[ht]
\vspace*{2cm}
\begin{center}
\leavevmode
\epsfbox{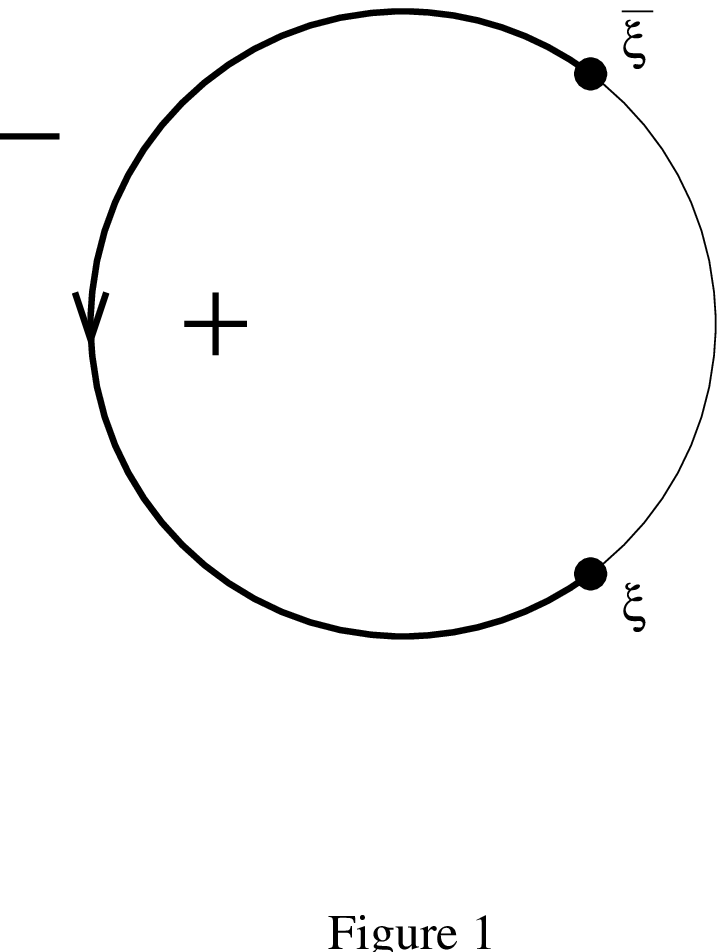}
\end{center}
\caption{\label{fig:contour}}
Contour $C$ on the unit circle for the integration with respect to $z$ in
the integral operator $\widehat{V}$. '$+$' and '$-$' indicate direction in
which the limit $z\to C$ has to be taken to obtain the boundary values
$\chi^\pm(z)$ for the Riemann-Hilbert problem in Section
\ref{sec:RH-problem}.
\end{figure}

Let us now consider the following infinite dimensional Riemann-Hilbert
problem for the integral operator-valued function $\chi(z)$:
\begin{enumerate}
\item
$\chi(z)$ is analytic outside the contour $C$ (Fig.~\ref{fig:contour}).

\item
$\chi^-(z) = \chi^+(z) \cdot L^{(m)}(z)$ for $z \in C$, and $\chi^\pm$ are
the boundary values of the function $\chi(z)$ as indicated in
Fig.~\ref{fig:contour}, and where the (integral-operator valued)
``conjugation matrix'' $L^{(m)}(z)$ is given by
\begin{equation}
   L^{(m)}(z) = I + \ell^{(m)}(z)
   \label{conjgmatrix}
\end{equation}
where $I(s,t) =\delta(s-t) \left(\begin{array}{cc} 1 & 0 \\ 0 & 1
\end{array} \right) $ and
\begin{equation}
   \ell^{(m)}(z|s,t) = \left( \begin{array}{cc}
      -e^{(m)}_+(z|s) e^{(m)}_-(z|t) & e^{(m)}_+(z|s) e^{(m)}_+(z|t) \\
      -e^{(m)}_-(z|s) e^{(m)}_-(z|t) & e^{(m)}_-(z|s) e^{(m)}_+(z|t)
      \end{array}
              \right)
\label{def:ell}
\end{equation}

\item
$\chi(z) \to I$ as $z\to\infty$.
\end{enumerate}

\noindent
In order to simplify our notations we will from now on suppress the
$m$-dependence in equations where all quantities are to be taken with
the same $m$ and write {\em e.g.} $\ell(z|s,t)$ instead of
$\ell^{(m)}(z|s,t)$.  In terms of the corresponding kernels the properties
1--3 can be rewritten in the following way:
\begin{itemize}
\item[P1.]
$\chi(z|s,t)$ is an analytic function of $z \not\in C$ for all $s,t$.

\item[P2.]
$\chi^-(z|s,t) = \chi^+(z|s,t) + \displaystyle{\int_0^\infty}
ds' \chi^+(z|s,s')\ \ell(z|s',t)$ for $z \in C$

\item[P3.]
$ \chi(z|s,t) = I(s,t)+{1\over z}\Psi_1(s,t)+\ldots$ as $z\to\infty$
\end{itemize}
\noindent
The connection of the RHP 1--3 to the IDE of Section \ref{sec:ide} is
summarized in the following lemma:\\
{\bf Lemma 1:} {\em Suppose now that the solution of the Riemann-Hilbert
problem 1--3 exists and is unique. Then the function $\Psi(z) = \chi(z)
\cdot \left(\begin{array}{cc} z^{-m} & 0 \\ 0 & 1 \end{array} \right)$
satisfies the integral operator-valued linear system (\ref{shift:fright}),
(\ref{f:dpsi2}).}
\vspace{0.5cm}

\noindent {\bf Proof : }
Applying the standard arguments based on Liouville's theorem and on
the $m$ independence of the conjugation integral operator $L_0(z)$
\begin{equation}
   L_0(z) = \left( \begin{array}{cc} z^{m} & 0 \\ 0 & 1 \end{array} \right)
       L(z) \left( \begin{array}{cc} z^{-m} & 0 \\ 0 & 1 \end{array} \right)
\end{equation}
we obtain
\begin{equation}
   \Psi^{(m+1)} \left[ \Psi^{(m)} \right]^{-1} =
   \left( \begin{array}{cc} 0 & 0 \\ 0 & 1 \end{array} \right) \cdot I\
   +\ {1\over z}\ U_0
   \label{Psi:shift}
\end{equation}
where $U_0 = \chi^{(m+1)} (0) \cdot \left( \begin{array}{cc} 1 & 0 \\
0 & 0 \end{array} \right) \cdot \left[ \chi^{(m)}(0) \right]^{-1}$.
As we will now show by determining $U_0$, (\ref{Psi:shift}) is the
integral-operator valued analog of (\ref{shift:fright}).

By construction the operator $\ell(z)$  (\ref{def:ell}) is nilpotent
$\ell^2(z)=0$, hence we have $L^{-1}(z) = I - \ell(z)$, which in turn
implies the equation
\begin{equation}
   \chi^{-1}(z|s,t) = \left(
   \begin{array}{cc} \chi_{22}(z|t,s) & -\chi_{12}(z|t,s) \\
                     -\chi_{21}(z|t,s) & \chi_{11}(z|t,s) \end{array}
   \right)
   \label{chi_inv}
\end{equation}
for the solution of the RH-Problem 1--3. Introducing notations
\begin{equation}
   \chi(0) = \left(
   \begin{array}{cc} 1-B_{+-} &   B_{++} \\
                      -B_{--} & 1+B_{-+} \end{array}
   \right)
   \label{chi0}
\end{equation}
we conclude from (\ref{chi_inv}) that
\begin{equation}
   \chi^{-1}(0) = \left(
   \begin{array}{cc} 1+B_{-+}^T &  -B_{++}^T \\
                       B_{--}^T & 1-B_{+-}^T \end{array}
   \right) \ .
   \label{chi0_inv}
\end{equation}

Substituting the expansion P3 for $\chi(z)$ as $z\to\infty$ into
(\ref{Psi:shift}), we see that $(U_0)_{11} = 1$. This together with
formulae (\ref{chi0}) and (\ref{chi0_inv}) allow us to rewrite $U_0$ as
\begin{equation}
  U_0 = \left( \begin{array}{cc}
  1 & - \left(1-B_{+-}^{(m+1)}\right)\left(B_{++}^{(m)}\right)^T \\[8pt]
  - B_{--}^{(m+1)}\left(1+\left(B_{-+}^{(m)}\right)^T\right)
             & B_{--}^{(m+1)} \left(B_{++}^{(m)}\right)^T
               \end{array} \right)\ .
\end{equation}
We now define functions $f_\pm(z|s)$ {\em via}
\begin{equation}
   \left( \begin{array}{c} f_+(z|s)\\f_-(z|s) \end{array} \right)
   = \int_0^\infty dt\ \chi^+(z|s,t)
   \left( \begin{array}{c} e_+(z|t)\\e_-(z|t) \end{array} \right)\ .
   \label{RH:rel}
\end{equation}
It can be shown analogously to Section XV.6 of \cite{vladb} and
\cite{itsx:90} that the functions $f_\pm$ defined this way are
identical to the ones defined in Section \ref{sec:ide} for
Eq.~(\ref{rkernelz}). This then implies that the kernels
$B_{ab}(s,t)$ from (\ref{chi0}) are just the kernels introduced
in (\ref{def:B}). To prove this one has to consider the canonical
integral representation for the solution of the problem 1--3
\begin{equation}
  \chi(z_1) = I - {1\over 2\pi i} \int_C {dz_2 \over z_2 - z_1}
           \chi^+(z_2)\ \ell(z_2)\ ,
\end{equation}
which implies
\begin{equation}
  \chi(0) = I + {i\over 2\pi} \int_C {dz \over z} \chi^+(z)\ \ell(z)\ .
\end{equation}
Taking into account the explicit formula for $\ell(z)$ and Eq.~(\ref{RH:rel})
we obtain the representations (\ref{def:B}) for $B_{ab}(s,t)$ defined in
(\ref{chi0}). This shows that $\Psi(z)$ fulfills the integral-operator
valued version of the system (\ref{shift:fright}), which can be
reobtained from (\ref{Psi:shift}) in the following way:
rewriting (\ref{Psi:shift}) as
\begin{equation}
   z^{-{1\over2}} \chi^{(m+1)}(z) \cdot \left(
     \begin{array}{cc} z^{-{1\over2}} & 0 \\ 0 & z^{1\over2}\end{array}
     \right) =
   \left[ \left( \begin{array}{cc} 0 & 0 \\ 0 & 1 \end{array} \right)
          \cdot I\ +\ {1\over z}\ U_0 \right] \chi^{(m)}(z)\ ,
   \label{chi:shift}
\end{equation}
and then applying both sides of (\ref{chi:shift}) to the vector $\left(
\begin{array}{c} e_+^{(m)}(z|s) \\ e_-^{(m)}(z|s) \end{array} \right)$ and
taking into account (\ref{RH:rel}) and that $e_\pm^{(m)} z^{\mp{1\over2}} =
e_\pm^{(m+1)}$ we arrive back at the equations (\ref{shift:fright}).
\vspace{0.5cm}

Let us now study the $\xi$-derivative of the function $\chi(z)$ in
order to connect (\ref{f:dpsi2}) to our RHP. The corresponding
analysis is very similar to the ones for the Bose gas \cite{itsx:90}
and the XXX chain \cite{frik:94}. In the neighbourhood of $C$ the function
$\chi(z)$ can be represented as
\begin{equation}
   \chi(z) = \widehat{\chi}(z) \cdot \chi_0(z)
   \label{def:chihat}
\end{equation}
where $\widehat{\chi}(z)$ is single-valued, invertible and analytic in that
neighbourhood, and
\begin{equation}
   \chi_0(z) = I -\ {i\over2\pi} \ln{z-\xib\over z-\xi}\ \ell(z)\ .
   \label{RH:chi0}
\end{equation}
{}From (\ref{def:chihat}) we conclude at once that
\begin{equation}
   {\partial \chi(z)\over \partial\psi} \cdot \chi^{-1}(z)
	= {A_+ \over z-\xib} + {A_- \over z-\xi}
   \label{RH:chipsi}
\end{equation}
where
\begin{equation}
  A_\pm = \lim_{z\to z_\pm} \left( z - z_\pm \right)
    \widehat{\chi}(z) \cdot {\partial \chi_0(z)\over \partial\psi} \cdot
     \chi_0^{-1}(z)  \cdot \widehat{\chi}^{-1}(z)\ , \qquad
      z_+ = \xib\ ,\quad z_- = \xi\ .
  \label{RH:defA}
\end{equation}
Differentiation of (\ref{RH:chi0}) gives
\begin{equation}
   \left( {\partial \chi_0(z)\over \partial\psi} \chi_0^{-1} \right)(z)
   = {1\over2\pi}\left[ {{\xib} \over z-\xib } + {{\xi} \over z-\xi}\right]\
   \ell(z)\ .
\label{RH:chi0psi}
\end{equation}
Using this together with $\ell^2=0$ and (\ref{def:ell}) and (\ref{RH:rel})
in (\ref{RH:defA}) we obtain
\begin{eqnarray}
   A_\pm(s,t) &=& {z_\pm \over 2\pi}
      \left(\widehat{\chi}\ \cdot\ \ell\ \cdot \widehat{\chi}^{-1}
      \right)(z_\pm|s,t)
     = {z_\pm \over 2\pi} \left(\chi\ \cdot\ \ell\ \cdot \chi^{-1}
                          \right)(z_\pm|s,t)
    \nonumber \\
   &=& {z_\pm \over 2\pi} \left(
   \begin{array}{cc}
       - f_+(z_\pm|s) f_-(z_\pm|t) & f_+(z_\pm|s) f_+(z_\pm|t) \\
       - f_-(z_\pm|s) f_-(z_\pm|t) & f_-(z_\pm|s) f_+(z_\pm|t)
   \end{array} \right)\ .
\end{eqnarray}
Recalling the definition of the potentials $F_\pm(s)$, $G_\pm(s)$, we can
rewrite the final formulae for $A_\pm$ as
\begin{eqnarray}
   A_+(s,t) &=& {\xib\over2\pi} \left(
   \begin{array}{cc}
       -G_+(s) G_-(t) & G_+(s)G_+(t) \\
       -G_-(s) G_-(t) & G_-(s)G_+(t)
   \end{array} \right)\ ,
   \nonumber \\
   && \label{RH:Afinal} \\
   A_-(s,t) &=& {\xi\over2\pi} \left(
   \begin{array}{cc}
       -F_+(s) F_-(t) & F_+(s)F_+(t) \\
       -F_-(s) F_-(t) & F_-(s)F_+(t)
   \end{array} \right)\ . \nonumber
\end{eqnarray}
{}From (\ref{RH:chipsi}) it follows that $\Psi(z)$ fulfills the
integral-operator valued version of (\ref{f:dpsi2})
\begin{equation}
   {\partial \Psi(z|s,t)\over \partial\psi}
	= \int_0^\infty dt' \left({A_+(s,t') \over z-\xib} +
          {A_-(s,t') \over z-\xi}\right)\Psi(z|t',t)\ .
 \label{psipsi}
\end{equation}
Acting with (\ref{psipsi}) on
$\left(\begin{array}{cc} f_+(z|t) \\ f_-(z|t)\end{array}\right)$ and
taking (\ref{RH:Afinal}) into account we arrive at the basic equation
(\ref{f:dpsi2}) for the $\psi$-derivative of $f_\pm(z|s)$. This
completes the proof of the Lemma.  {\bf q.e.d.}

\section{Asymptotics in some limiting cases
         \label{sec:special}}

In this section we obtain the large distance asymptotics of the FSFP
correlation functions in two limiting cases. The first case corresponds to
the limit of very strong magnetic fild. In the second case, the magnetic
field is arbitrary but the value of parameter $\eta$ is chosen to be
$3\pi/4$ corresponding to the free fermionic point.

For very large $h$ close to the critical field $h_c=4\cos^2\eta$ (for
which the ground state turns into the saturated ferromagnetic state) the
integration boundary $\Lambda$ in (\ref{intop}) tends to zero acoording to
\begin{equation}
   \Lambda = \frac{1}{2}|\tan\eta| \sqrt{h_c-h} + {\cal O}(h_c-h)\ .
\label{eq:Lambda-hc}
\end{equation}
In the limit $h\rightarrow h_c$, the kernel (\ref{vkernell}) can be
expanded to order ${\cal O}(m(h_c-h)^\frac{3}{2})$ as
\begin{equation}
   V^{(m)}(\lambda|\mu)\sim
   V_0(\lambda|\mu)=-\frac{1}{\pi}
	\frac{\sin(m|\cot\eta|(\lambda-\mu))}{\lambda-\mu}\ .
\label{eq:kernel-hc}
\end{equation}
We note that the dual fields do not contribute at all because
\begin{equation}
\phi(\l)=\phi(0)+{\cal O}((h_c-h)^{\frac{1}{2}}) ,
\end{equation}
and the dual expectation value can simply be dropped. Thus we find that in
the limit $h\to h_c$, $m\to\infty$ with $m(h_c-h)\ll 1\ $\footnote{This
inequality rather than $m(h_c-h)^\frac{3}{2}\ll 1$ is a consequence of a
more accurate analysis of the contribution of the terms dropped in
(\ref{eq:kernel-hc}) to the determinant.}
\begin{equation}
    P(m)\sim \det\left(1+V_0\right)\ .
\end{equation}
The long-distance asymptotics of this Fredholm determinant is known
\cite{dyson:76}
\begin{equation}
   \ln\det\left[1-\frac{1}{\pi}\frac{\sin(a(x-y))}{x-y}\right]\bigg|_{-s}^s
   \sim -\frac{1}{2}(as)^2 {\rm\ \ for}\ as\rightarrow\infty .
\end{equation}
In this way we obtain a Gaussian decay of the FSFP for large magnetic
fields $h\approx h_c$ at large distances
\begin{equation}
   P(m)\sim e^{-\frac{1}{8}\left( h_c-h \right) m^2}\quad
   {\rm for}\ \
   h\rightarrow h_c,\
   (h_c-h)^{-{1\over 2}} \ll m \ll (h_c-h)^{-1} \ .
\label{9}
\end{equation}

This result complements the expression for {\em near} asymptotics
obtained in \cite{efik:94}
\begin{equation}
   P(m) \sim e^{-{1\over \pi}\sqrt{h_c-h}\ m}\quad
   {\rm for}\ \
   h\rightarrow h_c,\
   m \ll (h_c-h)^{-{1\over 2}} \ .
\end{equation}
For the case of the XXX chain an analogous analysis can be performed
(based on the determinant representation \cite{korx:94}) giving the same
result (\ref{9}) (note that $h_c=4$ for the XXX case).

The above results (and more) can also be obtained from the Riemann-Hilbert
problem 1--3 presented at the beginning of Section \ref{sec:RH-problem}.
As shown above, in the limit $h\rightarrow h_c$ the dual fields can be
dropped and the integral kernel (\ref{vkernell}) can be approximated as
\begin{equation}
   V^{(m)}(\lambda|\mu) =\frac{i}{2\pi}\
	\frac{{\tilde e}_+(\lambda){\tilde e}_-(\mu)
               -{\tilde e}_-(\lambda){\tilde e}_+(\mu)}{\sinh(\lambda-\mu)}\ ,
   \qquad {\tilde e}_\pm(\lambda)=\left(\frac{\sinh(\lambda+i\eta)}
          {\sinh(\lambda-i\eta)}\right)^{\pm\frac{m}{2}}\ .
\label{vkernel_hc}
\end{equation}
It is worth emphasizing that this is the only approximation that we will
use in this approach. Performing the same change of variables as in
Sect.~\ref{sec:ide} this becomes an integral operator of the form (\ref{e})
with kernel
\begin{equation}
   V^{(m)}(z_1|z_2) = - {i\over{2\pi}} \
   {{e_+(z_1) e_-(z_2) - e_-(z_1) e_+(z_2)}\over{z_1-z_2}}, \qquad
   e_\pm(z) = z^{\mp{m\over2}}
\label{vkernel_hz}
\end{equation}
and the RHP 1--3 above reduces to the ordinary $2\times 2$ matrix RHP {\em
without $s$-integration}:
\begin{itemize}
\item[$\widetilde{\hbox{P1}}$.]
$\chi(z)$ is an analytic function of $z \not\in C$

\item[$\widetilde{\hbox{P2}}$.]
$\chi^-(z) = \chi^+(z)\pmatrix{0&z^{-m}\cr -z^m&2\cr}$

\item[$\widetilde{\hbox{P3}}$.]
$ \chi(\infty) = I .$
\end{itemize}
\noindent
The functional determinant of the operator (\ref{vkernel_hz}) can be
rewritten as the determinant of an $m\times m$ matrix from which the
behaviour for small $m$ is found. The large-$m$ asymptotics can be
obtained in terms of the RHP $\widetilde{\hbox{P1}}$--$\widetilde{\hbox{P3}}$
and yields the following asymptotic formula for
$\det(I+\widehat{V}^{(m)})$ \cite{dxi}
\begin{eqnarray}
   &&\ln\det\left(I+{\hat V}^{(m)}\right)\sim  m^2\ln\delta \quad
	\hbox{for}\ \ m\rightarrow\infty  \nonumber\\
   &&\delta= -\sin\frac{\psi}{2}\ ,\quad
   e^{i\psi}=\frac{e^{2\Lambda}w-1}{e^{2\Lambda}-w}\ ,
\label{16}
\end{eqnarray}
where $\Lambda\sim m^{-\frac{1}{2}-\epsilon}$. In terms of the magnetic
field this condition translates into $m(h_c-h)\sim m^{-2\epsilon}$. The
asymptotic equation (\ref{16}) is actually equivalent to (\ref{9}) as
\begin{eqnarray}
   \ln\delta &\sim& -\frac{1}{2}\ \Lambda^2 \cot^2\eta + O(\Lambda^4)\ ,
   \quad \hbox{for}\ \
   \Lambda\to 0\ \nonumber \\
   &\sim& -{1\over8}\ (h_c-h)\ , \quad \hbox{for}\ \
   h\to h_c \
\end{eqnarray}
Equations~(\ref{9}) and (\ref{16}) describe the long-distance
($m\rightarrow\infty$) asymptotics of $P(m)$ in the XXZ model in the
limit of small $\Lambda$.

Let us now consider the free fermionic point $\eta=3\pi/4$ (XX0 model) in
more detail. For this case the RHP
$\widetilde{\hbox{P1}}$--$\widetilde{\hbox{P3}}$ is the {\em exact} RHP and
(\ref{vkernel_hc}), (\ref{vkernel_hz}) are the {\em exact} integral
operators for the FSFP for {\em any} values of $\Lambda$. Thus (\ref{16})
with $\eta=3\pi/4$ is the result for the long-distance asymptotics of the
functional determinant of $I + \widehat{V}^{(m)}$ in the XX0 model for
arbitrary $\Lambda$ and thus arbitrary magnetic field $h$
\begin{equation}
   \ln\det\left(I+\widehat{V}^{(m)}\right)_{XX0}
	\sim  {m^2\over2}\ \ln\left({1\over2}+{1\over{2\cosh2\Lambda}}\right)
	 \quad
	\hbox{for}\ \ m\rightarrow\infty\  .
\end{equation}
For the XX0 model we have $\cosh2\Lambda = 2/h$. Hence we finally
obtain for the asymptotic behaviour of the FSFP at the free fermionic point
as a function of the magnetic field
\begin{equation}
	P_{XX0}(m) \sim \left({2+h\over4}\right)^{m^2\over2}
\end{equation}
which coincides with (\ref{9}) for $h\to h_c=2$.

On the basis of the above results we conjecture that the FSFP exhibits a
Gaussian decay for any value of $\eta$ and any value of the magnetic field
$h<h_c$.

\section*{Acknowledgements}
We are grateful to the Aspen Center for Physics, where much of this work
was performed.  F.H.L.E. is grateful to the Institute for Theoretical
Physics in Stony Brook for hospitality. This work was partially supported
by the Deutsche For\-schungs\-gemein\-schaft under Grant No.\ Fr~737/2--1
and by the National Science Foundation (NSF) under Grant Nos.\ DMS-9315964
and PHY-9321165.





\end{document}